\documentclass[prd,showpacs,preprintnumbers,superscriptaddress,floatfix,twocolumn,amsmath,amssymb]{revtex4}
\usepackage{epsfig}
\usepackage{graphicx}
\usepackage{dcolumn}
\usepackage{bm}
\usepackage[active]{srcltx}
\newcommand{\be}{\begin{equation}}

\newcommand{\ee}{\end{equation}}
\newcommand{\bea}{\begin{eqnarray}}
\newcommand{\eea}{\end{eqnarray}}

\begin{document}


\title{Abnormal effective connectivity in migraine with aura under photic stimulation}
\date{\today}
\author{
S. Stramaglia}\affiliation{Dipartimento di Fisica, Universit\'a
degli Studi di Bari and INFN, via Orabona 4, 70126 Bari,
Italy\\}\author{D. Marinazzo}\affiliation{Faculty of Psychology and
Educational Sciences, Department of Data Analysis, Ghent University,
Henri Dunantlaan 1, B-9000 Gent, Belgium\\}\author{M.
Pellicoro}\affiliation{Dipartimento di Fisica, Universit\'a degli
Studi di Bari and INFN, via Orabona 4, 70126 Bari,
Italy\\}\author{M. de Tommaso} \affiliation{Dipartimento di Scienze
Neurologiche e Psichiatriche, Universit\'a degli Studi di Bari,
Piazza Giulio Cesare 11, 70124 Bari, Italy\\}

\date{\today}

\begin{abstract}
Migraine patients with aura show a peculiar pattern of visual
reactivity compared with those of migraine patients without aura: an
increased effective connectivity, connected to a reduced
synchronization among EEG channels, for frequencies in the beta
band. The effective connectivity is evaluated in terms of the
Granger causality. This anomalous response to visual stimuli may
play a crucial role in the progression of spreading depression and
clinical evidences of aura symptoms.

\pacs{87.19.L-, 05.45.Tp, 05.45.Xt, 42.66.Lc}
\end{abstract}

\maketitle Important information on the structure of complex systems
can be obtained by measuring to what extent the individual
components exchange information among each other. Transfer entropy
\cite{schreiber,lehnertz} and Granger causality
\cite{granger,bli,ding,kc} have emerged in recent years as leading
statistical techniques to detect cause-effect relationships between
time series; they are equivalent in the case of Gaussian stochastic
variables \cite{seth}. These approaches provide further insights on
the architecture of complex systems, in addition to those from
correlation and synchronization analysis
\cite{boccaletti,rosenblum,quiroga}. In neuroscience,
interdependencies estimated by correlation or spectral coherence are
referred to as functional connectivity, whilst effective
connectivity is a notion related to Granger causality or transfer
entropy \cite{friston}; the relationship between functional and
effective connectivity in the cortex represents a significant
challenge to present-day neuroscience
\cite{sporns,noineuroimage,neuro,fr}.
\begin{figure}[ht]
\includegraphics[width=8.5cm]{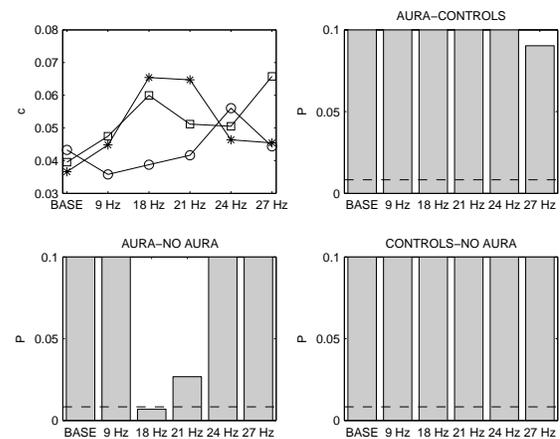}\caption{{\rm
(Top left) The linear Granger causality of EEGs filtered in the beta
band, averaged over pairs of channels and over subjects in the
classes, is depicted for aura patients (stars), no aura patients
(empty circles) and controls (empty squares) for basal condition and
as a function of the frequency of stimulations. We also make the
supervised analysis (hypothesis testing) of the linear Granger
causality, averaged over pairs of channels, in the beta band. The
probabilities that the measured values from two classes were drawn
from the same distribution, evaluated by {\it t} test, are depicted:
aura vs controls (top right) aura vs no aura (bottom left) and aura
vs controls (bottom right). Dashed lines correspond to the
significance threshold at 5$\%$ after Bonferroni's correction.
\label{fig0}}}\end{figure}

Migraine, an incapacitating disorder of neurovascular origin,
consists of attacks of headache, accompanied by autonomic and
possibly neurological symptoms \cite{class}. This pathology affects
a relevant fraction of the general population and represents a
social problem. The study of phase synchronization in EEG rhythms
showed a pattern of alpha rhythm (8-12.5 Hz) hyper-synchronization
under repetitive flash stimulation in migraine without aura
patients, opposite to a de-synchronization trend in non-migraine
subjects \cite{noiprl}; this nonlinear EEG pattern was found to be
modulated by anti-epileptic drugs \cite{drugs}. Approximately
one-third of people who suffer migraine headaches perceive an aura
(visual abnormality lasting 10-30 minutes) as a sign that the
migraine will soon occur \cite{aureview}. There are evidences
\cite{sp} of relationships between migraine aura and the spreading
depression (SD) phenomenon (a wave of electrophysiological
hyperactivity followed by a wave of inhibition \cite{leao}) but the
link between SD and headache is far to be explained in humans
\cite{zhang}. SD in clinically less conspicuous or extended brain
regions may be the trigger of migraine attacks ostensibly without a
"perceived" aura \cite{ayata}, so it may be plausible that, migraine
with and without aura may differ for those neuronal factors favoring
the clinical expression of SD phenomenon.

\begin{figure}[ht]
\includegraphics[width=8.5cm]{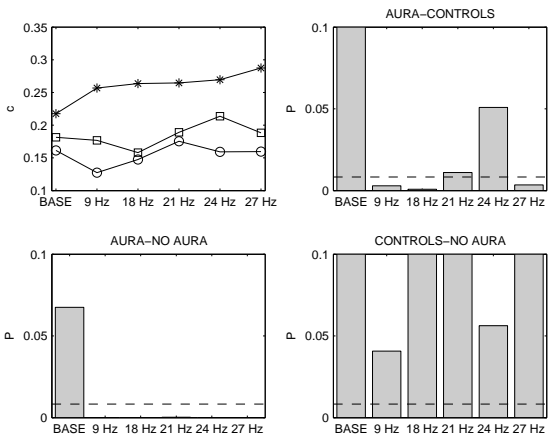}\caption{{\rm
(Top left) The nonlinear Granger causality of EEGs filtered in the
beta band, averaged over pairs of channels and over subjects in the
classes, is depicted for aura patients (stars), no aura patients
(empty circles) and controls (empty squares) for basal condition and
as a function of the frequency of stimulations. A Gaussian kernel
with $\sigma =10$ is used. We also make the supervised analysis
(hypothesis testing) of the nonlinear Granger causality, averaged
over pairs of channels, in the beta band. The probabilities that the
measured values from two classes were drawn from the same
distribution, evaluated by {\it t} test, are depicted: aura vs
controls (top right) aura vs no aura (bottom left) and aura vs
controls (bottom right). Dashed lines correspond to the significance
threshold at 5$\%$ after Bonferroni's correction. The ANOVA
\cite{anova}, taking the nonlinear Granger causality as variable,
the diagnosis (migraine with aura vs migraine without aura vs
controls) and the frequency of stimulation (9-18-21-24-27 Hz) as
factors, yields the F values 377.35, 8.36 and 3.38 for diagnosis,
frequency and interaction respectively; all the three factors are
thus recognized as significant.
 \label{fig1}}}\end{figure}

The question we address here is the following: how does the response
of aura migraine patients, to external stimuli, differ from those of
patients without aura? We show that, in presence of visual
stimulations, migraineurs with aura show a pattern of increased
effective connectivity between EEG channels in the beta band, and
correspondingly a decrease of the synchronization between channels.
This variation of connectivity, due to stimuli, is statistically
significant for migraine with aura: we remark that in our knowledge
neurophysiological patterns separating migraine with and without
aura have been rarely detected \cite{conte},  nor relevant
differences in regard to visual reactivity \cite{genco}.

\begin{figure}[ht]
\includegraphics[width=8.5cm]{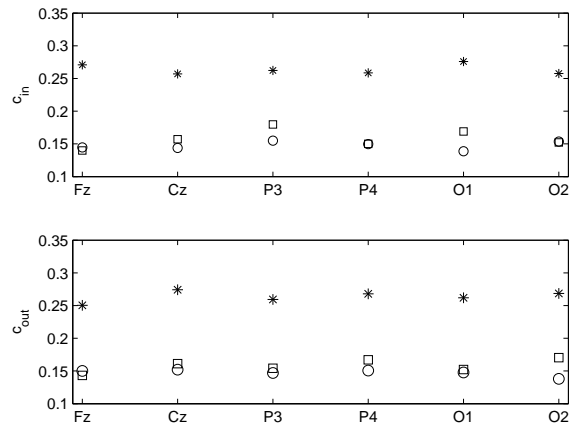}\caption{{\rm
For each electrode, the incoming (top) and outgoing (bottom)
Gaussian kernel Granger causality in the beta band, corresponding to
18 Hz stimuli and averaged over subjects in classes, is depicted;
aura patients (stars), no aura patients (empty circles) and controls
(empty squares).
 \label{fig2}}}\end{figure}

Our data are as follows. EEG is recorded from 19 patients (7 males,
20-44 age) affected by migraine with aura, 19 (4 males, 21-45 age)
patients affected by migraine without aura, and from 11 healthy
subjects (control group, 3 males, 20-46 age). All patients are in
the interictal state, the time from the end of the last attack being
at least 72 h. No patient was under preventive treatment nor had
assumed symptomatic drugs in the 72 hours preceding the recording
session. During the acquisition, flash stimuli are presented at a
rate of 9-18-21-24-27 Hz; also EEG in the absence of stimuli (base)
is recorded. Each frequency of stimulation is delivered by a flash
with 0.2 J luminance for about 20 sec. EEG data are recorded by six
scalp electrodes: two occipital channels (O1 and O2), two parietal
ones (P3 and P4), a central electrode (Cz) and a frontal one (Fz);
the sampling rate is 256 Hz, and the EEG is digitally filtered off
line by a filter with a band-pass 0.3-30 Hz.

\begin{figure}[ht]
\includegraphics[width=8.5cm]{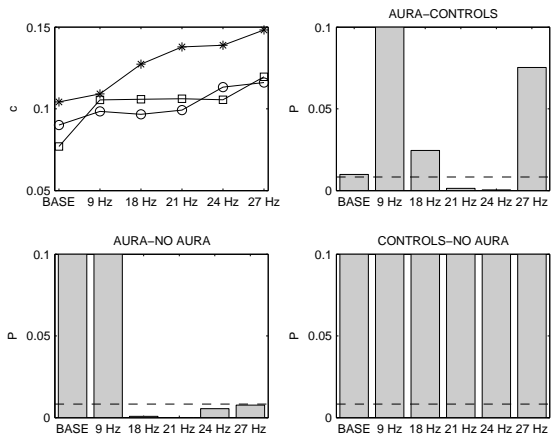}\caption{{\rm
(Top left) The nonlinear Granger causality of EEGs filtered in the
beta band, averaged over pairs of channels and over subjects in the
classes, is depicted for aura patients (stars), no aura patients
(empty circles) and controls (empty squares) for basal condition and
as a function of the frequency of stimulations. The polynomial
kernel with $p=2$ is used. We also make the supervised analysis
(hypothesis testing) of the polynomial Granger causality, averaged
over pairs of channels, in the beta band. The probabilities that the
measured values from two classes were drawn from the same
distribution, evaluated by {\it t} test, are depicted: aura vs
controls (top right) aura vs no aura (bottom left) and aura vs
controls (bottom right). Dashed lines correspond to the significance
threshold at 5$\%$ after Bonferroni's correction.
 \label{fig3}}}\end{figure}

Next, we describe our findings. Firstly we investigate the alpha
band hyper-synchronization phenomenon \cite{noiprl} in presence of
flash stimuli. The pattern of \cite{noiprl} is confirmed for
migraineurs without aura, whilst patients with aura do not show
alpha band hyper-synchronization in presence of light stimuli: in
this range of frequencies, aura patients seem to behave as controls.

\begin{figure}[ht]
\includegraphics[width=8.5cm]{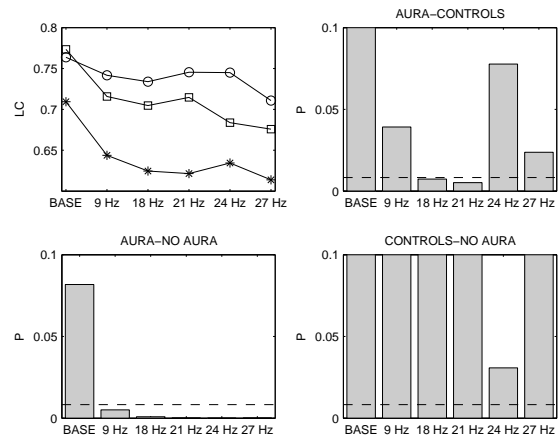}\caption{{\rm
(Top left) The linear correlation of EEGs filtered in the beta band,
averaged over pairs of channels and over subjects in the classes, is
depicted for aura patients (stars), no aura patients (empty circles)
and controls (empty squares) for basal condition and as a function
of the frequency of stimulations. We also make the supervised
analysis (hypothesis testing) of the linear correlations, averaged
over pairs of channels, in the beta band. The probabilities that the
measured values from two classes were drawn from the same
distribution, evaluated by {\it t} test, are depicted: aura vs
controls (top right) aura vs no aura (bottom left) and aura vs
controls (bottom right). Dashed lines correspond to the significance
threshold at 5$\%$ after Bonferroni's correction. The ANOVA
\cite{anova}, taking the linear correlation as variable, the
diagnosis (migraine with aura vs migraine without aura vs controls)
and the frequency of stimulation (9-18-21-24-27 Hz) as factors,
yields the F values 143.42, 3.67 and 0.74 for diagnosis, frequency
and interaction respectively; diagnosis and frequency of stimulation
are recognized as significant factors. \label{fig4}}}\end{figure}

Moving to higher frequencies (beta band, 12.5-30 Hz), instead, we
find a peculiar pattern of visual reactivity for aura patients. We
set the order of the regression model equal to $m=6$ \cite{bic} and
evaluate the linear Granger causality in the beta band \cite{seth1}:
linear causalities are weak and show significant differences among
classes only at 18 Hz stimulations, see figure (\ref{fig0}). On the
other hand, evaluating the nonlinear Granger causality among the
filtered EEG signals with the kernel approach described in \cite{kc}
and a Gaussian kernel, we find that aura migraine patients exhibit
increased values of causality in presence of stimuli, whilst
controls and no aura patients do not show significant variation
w.r.t. basal conditions. In particular the discrimination between
aura and no-aura patients is excellent in presence of flash stimuli.
Figure (\ref{fig1}) refers to $\sigma =10$ as the width of the
Gaussian kernel, however the results are robust to variations of
$\sigma$.

\begin{figure}[ht]
\includegraphics[width=8.5cm]{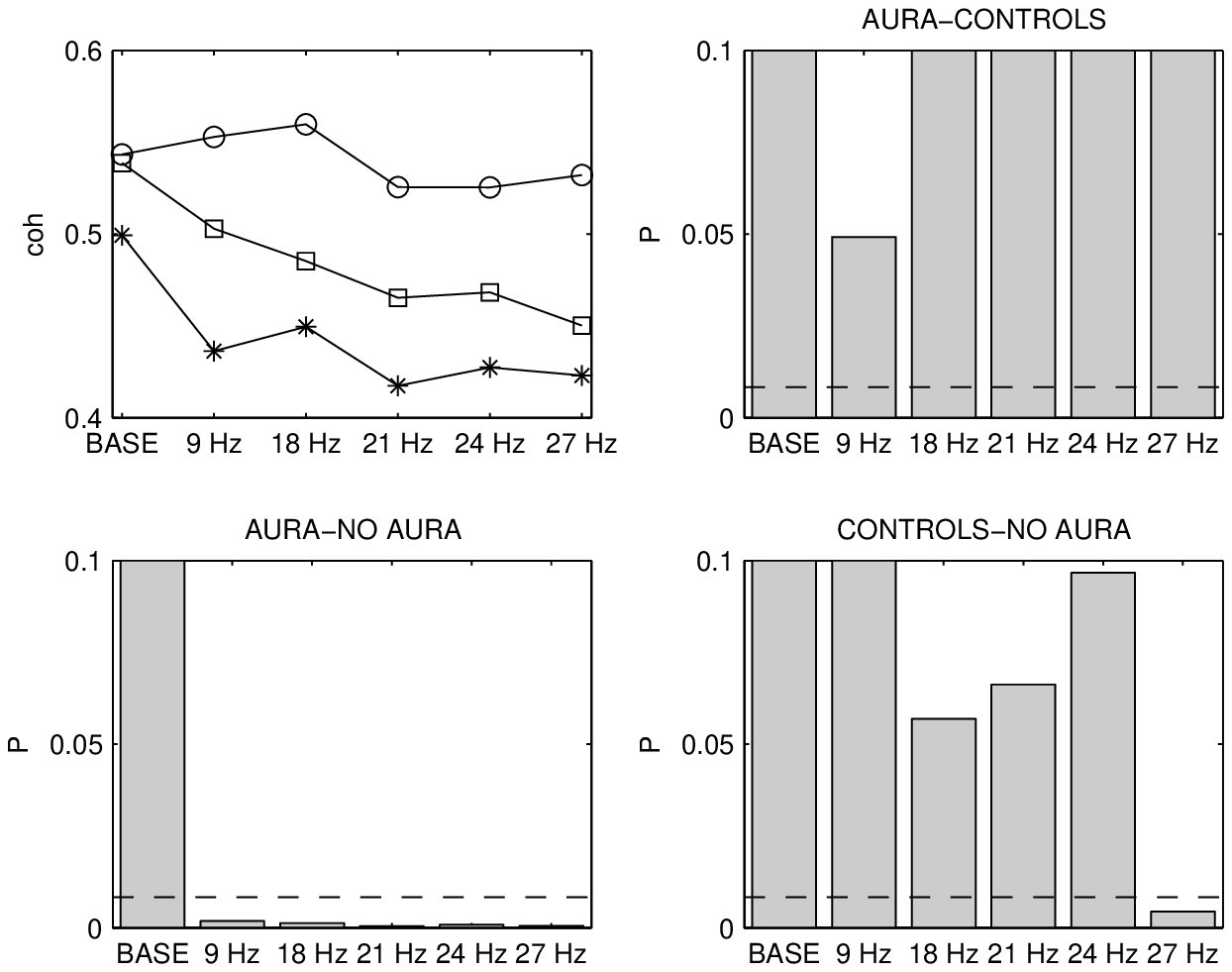}\caption{{\rm
(Top left) The coherence function of EEGs in the beta band, averaged
over pairs of channels and over subjects in the classes, is depicted
for aura patients (stars), no aura patients (empty circles) and
controls (empty squares) for basal condition and as a function of
the frequency of stimulations. We also make the supervised analysis
(hypothesis testing) of the coherence, averaged over pairs of
channels, in the beta band. The probabilities that the measured
values from two classes were drawn from the same distribution,
evaluated by {\it t} test, are depicted: aura vs controls (top
right) aura vs no aura (bottom left) and aura vs controls (bottom
right). Dashed lines correspond to the significance threshold at
5$\%$ after Bonferroni's correction.
 \label{fig5}}}\end{figure}

A topographic analysis is also performed: for each electrode, we
evaluate the total incoming causality (the sum of the causalities
from the other electrodes to the electrode under consideration) as
well as the total outgoing causality (the sum of the causalities
from the electrode under consideration to the other electrodes). In
figure (\ref{fig2}) we depict these quantities for the three classes
at 18 Hz stimulations in the beta band: it shows that the phenomenon
is diffuse over the scalp.

Use of the Gaussian kernel ensures that all orders of nonlinearities
are taken into account. If one considers only nonlinearities up to
the second order, weaker causalities are detected and less
discriminating power is obtained, see figure (\ref{fig3}) where we
describe the application of nonlinear Granger causality with a
polynomial kernel of degree two. We conclude that a relevant amount
of nonlinear information transmission characterizes this phenomenon.


Turning to synchronization, we  consider the Pearson linear
correlation between channels and find that the increased flow of
information, due to flash stimuli, is connected to weakening of the
correlations between them. In figure (\ref{fig4}) we depict the
linear correlation between signals filtered in the beta band,
averaged over all pairs of channels; in presence of light
stimulations the strength of correlations decreases for aura
patients. We find similar results also in terms of the coherence
function averaged in the beta band, see figure (\ref{fig5}), as well
as for the beta band phase synchronization \cite{tass}.

We also quantify the separation among classes in terms of the ROC
area \cite{roc}, which is directly related to the separation of two
conditional distributions, and measures the discrimination ability
of the forecast. For any frequency of stimulation, using the
Gaussian kernel Granger causality we obtain a roc area equal to 0.87
for aura vs no-aura patients. A similar result is found using the
linear correlation, roc area equal to 0.82 for aura vs no-aura at
all frequencies.

Summarizing, we have described for the first time a
neurophysiological pattern which seems peculiar of migraine patients
perceiving visual aura, where the loss of synchronization between
channels, due to light, induces stronger statistical causal
connections among them, diffuse over the scalp. This pattern is
characterized by nonlinear transfer of information among channels
and provides excellent discrimination between aura and no-aura
patients. The biological implications of this complex phenomenon in
facilitating SD progression and aura symptoms perception is the
challenge for a better  understanding of migraine pathophysiology.

\end{document}